\documentclass[12pt,english,floatfix,superscriptaddress,aps,prd,preprint]{revtex4}
\usepackage{amsmath}
\usepackage{amssymb}
\usepackage{amsbsy}
\usepackage{amsfonts}
\usepackage{amsopn}
\usepackage{amstext}
\usepackage{graphicx}
\usepackage{amssymb}
\usepackage{amsfonts}
\usepackage{amsmath}
\usepackage{graphicx}
\usepackage{color}
\usepackage{slashed}
\usepackage{esint}
\usepackage[dvips]{epsfig}
\usepackage[dvips]{graphicx}
\usepackage{float}
\usepackage{units}
\usepackage{textcomp}
\newcommand{\beq}{\begin{equation}}
\newcommand{\eeq}{\end{equation}}
\newcommand{\bea}{\begin{eqnarray}}
\newcommand{\eea}{\end{eqnarray}}

\begin{document}

\title{Generalized Ellis-Bronikov traversable wormholes in $f(R)$ gravity with anisotropic dark matter}

\author{C. R. Muniz\footnote{E-mail: celio.muniz@uece.br}}\affiliation{Universidade Estadual do Cear\'a, Faculdade de Educa\c c\~ao, Ci\^encias e Letras de Iguatu, 63500-000, Iguatu, CE, Brazil.}
\author{R. V. Maluf\footnote{E-mail: rvmaluf@fisica.ufc.br}}\affiliation{Universidade Federal do Cear\'{a}, Centro de Ci\^encias, Departamento de F\'{i}sica 60000-000, Fortaleza, CE, Brazil.}

\date{\today}

\begin{abstract}
This paper studies generalized Ellis-Bronikov (E-B) traversable wormholes in $f(R)$ extended gravity. We assume that these wormholes are supported by anisotropic dark matter (DM) according to the most often used phenomenological models, namely those of Navarro-Frenk-White (NFW), Thomas-Fermi (T-F), and Pseudo-isothermal (PI). Initially, we obtain the field equations in a general scenario of $f(R)$ theories in the metric formalism for the static and spherically symmetric Morris-Thorne spacetime. Then we particularize to a $f(R)$ model with power-law, including the Starobinky modified gravity. Following, we analyze the energy conditions which are not dependent on the DM models (Null and Weak Energy Conditions -- NEC and WEC) and those which are model-dependent (Strong and Dominant Energy Conditions SEC and DEC). Finally, we compare some E-B wormhole solutions and the mentioned DM models, discussing the feasibility of the wormhole-dark matter system in different scenarios.
\end{abstract}


\maketitle
\section{Introduction}

The $\Lambda$CDM model of the standard cosmology, supported by observational data, preconizes that the stuff of our universe consists of about 4$\%$ baryonic matter, 29.6$\%$ dark matter, and 67.4$\%$ dark energy \cite{Planck:2018vyg}. Concerning galaxies and their clusters, dark matter strongly contributes to the formation, evolution, and coalescence of these structures \cite{Trujillo-Gomez:2010jbn} through the only fundamental interaction that it is apparently capable of experiencing: gravity.

Dark matter halos are present in the vast majority of galaxies, and thus one has considered the formation of traversable wormholes inside them (see, e.g., \cite{Sarkar:2019uhk,Jusufi:2019knb,Xu:2020wfm}). These hypothetical objects are predicted by general relativity (GR) and represent a kind of tunnel in the spacetime that connects two distant regions of the same universe or two different universes (see \cite{Morris:1988cz,Visser}, and references therein). This subject has been recently investigated in a more fundamental level from the J. Maldacena works \cite{Maldacena:2013xja,Maldacena:2017axo,Maldacena:2020sxe} and also applied in the context of condensed matter systems \cite{Gonzalez:2009je,Alencar:2021ejd}. Usually, it is required some type of exotic matter sourcing traversable wormholes. However, in scenarios of modified theories of gravity, such a feature can change with non-exotic matter working as a source for the wormhole geometry  \cite{Pavlovic:2014gba,Mehdizadeh:2019qvc,Sahoo:2020sva,Moti,Alencar,Sadeghi:2022sto}.

Modified $f(R)$ gravity has also drawn much attention in the last few years. It is based on a generalization of Einstein's field equations by replacing the Ricci scalar curvature, $R$, with a general function, $f(R)$, in the gravitational action. A relevant feature of $f(R)$ gravity is that, differently from $\Lambda$CDM standard cosmology based on GR, it is unnecessary postulating dark energy or introduce any kind of new matter field to explain the cosmic inflation \cite{Kehagias:2013mya,Ketov,Aziz:2021evx,Sharma:2022tce} and the present phase of the Universe accelerated expansion \cite{Nojiri,Sotiriou}. In fact, the simplest $f(R)$ model, due to A. Starobinsky \cite{Starobinsky1,Starobinsky2}, with an additional Ricci scalar quadratic term in the Einstein-Hilbert action and therefore only one free parameter, leads to a de Sitter phase for as long as this term dominates, being compatible with the experimental data collected from the Planck satellite \cite{Akrami}.

In this paper, we will investigate the possibility of occurring generalized Ellis-Bronikov (E-B) traversable wormholes in $f(R)$ extended gravity. The wormholes are admitted to be sourced by anisotropic dark matter compatible with the most often used phenomenological models, namely those of Navarro-Frenk-White (NFW), Thomas-Fermi (T-F) and pseudo-isothermal (PI). Thus, we will obtain the general field equations of any $f(R)$ theory in the metric formalism, and then we will particularize to a power-law $f(R)$ model, including the Starobinky modified gravity. Following, we will analyze the energy conditions associated with the source, which are not dependent on the DM models (Null and Weak Energy Conditions -- NEC and WEC) and those that are model-dependent (Strong and Dominant Energy Conditions -- SEC and DEC). Finally, we compare some E-B wormhole solutions with the mentioned DM profiles and the $f(R)$ Starobinsky-like models regarding the energetic feasibility of wormhole-dark matter systems.

The paper is organized as follows: Section \ref{rev} reviews the general Morris-Thorne wormhole solution in $f(R)$ gravity. Section \ref{DM} discusses the generalized E-B wormhole sourced by anisotropic dark matter in the Starobinsky $f(R)$ model. In section \ref{EC}, we analyze the energy conditions of the system wormhole-dark matter, and then finally present the conclusions and close the paper in section \ref{Con}.

\section{Wormhole solutions in $f(R)$ gravity\label{rev}}
Let us start by considering the following action for a $f(R)$ modified theories of gravity:
\begin{equation}\label{action1}
S=\frac{1}{2\kappa}\int d^4x\sqrt{-g}\;f(R)+S_M(g^{\mu\nu},\psi)
\,,
\end{equation}
where $\kappa =8\pi G$ will be taken equal to the unit $(\kappa=1$) in order to keep simplicity and $f(R)$ is initially an arbitrary function of the curvature scalar. Besides this, $S_M(g^{\mu\nu},\psi)$ is
the matter action with $S_M=\int d^4x\sqrt{-g}\;{\cal
L}_m(g_{\mu\nu},\psi)$, where ${\cal L}_m$ is the matter
Lagrangian density, in which the matter is minimally coupled to the
metric $g_{\mu\nu}$ and $\psi$ denotes the matter
fields.

Varying the action (\ref{action1}) with respect to the metric, we get the gravitational field equation:
\begin{equation}
FR_{\mu\nu}-\frac{1}{2}f\,g_{\mu\nu}-\nabla_\mu \nabla_\nu
F+g_{\mu\nu}\Box F=\,T^{(m)}_{\mu\nu} \,,
    \label{field:eq}
\end{equation}
where we define $F\equiv df/dR$ and $\Box=g^{\alpha\beta}\nabla_{\alpha}\nabla_{\beta}$. The energy-momentum tensor associated with the matter content is given by
\begin{equation}
T^{(m)}_{\mu\nu}=-\frac{2}{\sqrt{-g}}\frac{\delta\sqrt{-g}{\cal L}_m}{\delta g^{\mu\nu}}.
\end{equation}
Considering the contraction of Eq.
(\ref{field:eq}), we arrive to the following relation
\begin{equation}
FR-2f+3\,\Box F=\,T \,,
 \label{trace}
\end{equation}
which shows that the Ricci scalar is a fully dynamical degree of
freedom, and $T=T^{(m)\mu}{}_{\mu}$ is the trace of the energy-momentum
tensor.

The trace equation (\ref{trace}) can be used to simplify the field
equations and then can be kept as a constraint equation. Thus,
substituting the trace equation into Eq. (\ref{field:eq}) and
re-organizing the terms, we end up with the following gravitational
field equation
\begin{equation}
G_{\mu\nu}\equiv R_{\mu\nu}-\frac{1}{2}R\,g_{\mu\nu}= T^{{\rm
eff}}_{\mu\nu} \,,
    \label{field:eq2}
\end{equation}
where the effective stress-energy tensor is given by $T^{{\rm
eff}}_{\mu\nu}= T^{(c)}_{\mu\nu}+\tilde{T}^{(m)}_{\mu\nu}$. The
term $\tilde{T}^{(m)}_{\mu\nu}$ corresponds to
\begin{equation}
\tilde{T}^{(m)}_{\mu\nu}=T^{(m)}_{\mu\nu}/F \,,
\end{equation}
and the curvature stress-energy tensor, $T^{(c)}_{\mu\nu}$, is
defined as
\begin{eqnarray}
T^{(c)}_{\mu\nu}=\frac{1}{F}\left[\nabla_\mu \nabla_\nu F
-\frac{1}{4}g_{\mu\nu}\left(RF+\Box F+T\right) \right]    \,.
    \label{gravfluid}
\end{eqnarray}

It is also interesting to consider the conservation law for the
above curvature stress-energy tensor. Taking into account the
Bianchi identities, $\nabla^\mu G_{\mu\nu}=0$, and the
diffeomorphism invariance of the matter part of the action, which
yields $\nabla^\mu T^{(m)}_{\mu\nu}=0$, we verify that the
effective Einstein field equation provides the following
conservation law
\begin{equation}\label{conserv-law}
\nabla^\mu T^{(c)}_{\mu\nu}=\frac{1}{F^2}
T^{(m)}_{\mu\nu}\nabla^\mu F  \,.
\end{equation}

Once the scenario for the $f(R)$ theories of gravity has been presented, we will study the wormhole solutions in this framework. The following line element describes the geometry of a static and spherically symmetric traversable wormholes \cite{Morris:1988cz}:
\begin{equation}
ds^2=-e^{2\Phi(r)}dt^2+\frac{dr^2}{1-b(r)/r}+r^2\,(d\theta^2 +\sin
^2{\theta} \, d\phi ^2) \,,
    \label{metric}
\end{equation}
where $\Phi(r)$ and $b(r)$ are arbitrary functions of the radial
coordinate, $r$, denoted as the redshift function, and the shape
function, respectively. The radial coordinate $r$ decreases from infinity to a minimum value $r_0$, the radius of the throat, where $b(r_0)=r_0$. A fundamental property of a wormhole is that a flaring out condition of the throat, given by $(b-b^{\prime}r)/b^{2}>0$ \cite{Morris:1988cz}, and at the throat
$b(r_{0})=r=r_{0}$, such that the condition $b^{\prime}(r_{0})<1$ is imposed
to have wormhole solutions. It is precisely these restrictions
that impose the NEC violation in classical general relativity.
Another condition that needs to be satisfied is $1-b(r)/r>0$. For
the wormhole to be traversable, one must demand that there are no
horizons present, which are identified as the surfaces with
$e^{2\Phi}\rightarrow0$, so that $\Phi(r)$ must be finite
everywhere. In the analysis outlined below, we consider that the
redshift function is constant, $\Phi'=0$, which simplifies the
calculations considerably, and provide interesting exact wormhole
solutions.

Relative to the matter content of the wormhole, we impose that the
energy-momentum tensor that threads the wormhole satisfies the
energy conditions, which will be discussed in section \ref{EC}, and is given by the following anisotropic distribution of matter
\begin{equation}
T^{(m)}_{\mu\nu}=(\rho+p_t)U_\mu \, U_\nu+p_t\,
g_{\mu\nu}+(p_r-p_t)\chi_\mu \chi_\nu \,,
\end{equation}
where $U^\mu$ is the four-velocity, $\chi^\mu$ is the unit
spacelike vector in the radial direction, i.e.,
$\chi^\mu=\sqrt{1-b(r)/r}\,\delta^\mu{}_r$. $\rho(r)$ is the
energy density, $p_r(r)$ is the radial pressure measured in the
direction of $\chi^\mu$, and $p_t(r)$ is the transverse pressure
measured in the orthogonal direction to $\chi^\mu$. Taking into
account the above considerations, the stress-energy tensor is
given by the following profile: $T^{(m)\mu}{}_{\nu}={\rm
diag}[-\rho(r),p_r(r),p_t(r),p_t(r)]$.

Taking into account the above considerations, the effective field equation (\ref{field:eq2}) provides the
following equations for the wormhole geometry
\begin{eqnarray}
-\frac{b'}{r^{2}}+\frac{\rho}{F}+\frac{H}{F}	&=&0,\label{fieldtt}\\
-\frac{b}{r^{3}}-\frac{p_{r}}{F}+\frac{1}{F}\left[H+F''\left(\frac{b}{r}-1\right)+F'\left(\frac{b'}{2r}-\frac{b}{2r^{2}}\right)\right]	&=&0,\label{fieldrr}\\
-\frac{b'}{2r^{2}}+\frac{b}{2r^{3}}-\frac{p_{t}}{F}+\frac{1}{F}\left[H+F'\left(\frac{b}{r^{2}}-\frac{1}{r}\right)\right]	&=&0.\label{fieldthetatheta}
\end{eqnarray}where the prime denotes a derivative with respect to the radial
coordinate, $r$. The term $H=H(r)$ is defined as
\begin{equation}
H(r)=\frac{1}{4}\left(FR+\Box F +T\right) \,,
\end{equation}
and the Ricci curvature scalar is $R=2b'(r)/r^{2}$. Thus, the term $\Box F$ is encoded by the following expression
\begin{equation}
\Box F=F''\left(1-\frac{b}{r}\right)-F'\left(\frac{b'}{2r}+\frac{3b}{2r^{2}}-\frac{2}{r}\right).
\end{equation}So we can explicitly write $H(r)$ as
\begin{equation}
H=-\frac{\rho}{4}+\frac{p_{r}}{4}+\frac{p_{t}}{2}+\frac{Fb'}{2r^{2}}-F'\left(\frac{b'}{8r}+\frac{3b}{8r^{2}}-\frac{1}{2r}\right)-F''\left(\frac{b}{4r}-\frac{1}{4}\right).\label{H}
\end{equation}

Note that the field equations (\ref{fieldtt})-(\ref{fieldthetatheta}), despite their complexities, form an algebraic system of equations for the components of the energy-momentum tensor. From the field equations (\ref{fieldtt})-(\ref{fieldthetatheta}), and with the help of (\ref{H}), we can express the radial and transverse pressures in the following way
\begin{eqnarray}\label{Pe}
p_{r}&=&-\rho+F\left(\frac{b'}{r^{2}}-\frac{b}{r^{3}}\right)+F'\left(\frac{b'}{2r}-\frac{b}{2r^{2}}\right)+F''\left(\frac{b}{r}-1\right),\\\label{pezinho}
p_{t}&=&-\rho+F\left(\frac{b'}{2r^{2}}+\frac{b}{2r^{3}}\right)+F'\left(\frac{b}{r^{2}}-\frac{1}{r}\right).
\end{eqnarray}

It is worth mentioning that the solutions obtained for $p_{r}$ and $p_{t}$ depend on the energy density $\rho$ and on the explicit forms of the functions $F(r)$ and $b(r)$. In this work, we are interested in studying the behavior of these pressures and evaluating the resulting energy conditions in a dark matter context considering the generalized Ellis-Bronikov model for the form of the wormhole function $b(r)$. We will also assume the Starobinsky-like model for the $f(R)$ function, motivated by its phenomenological feasibility in the cosmological context. Such models will be better exemplified in what follows.

\section{Generalized Ellis-Bronikov wormhole sourced by anisotropic dark matter in the Starobinsky-like model\label{DM}}
The generalised Ellis--Bronnikov model of wormholes was discussed in \cite{Kar:1995jz,DuttaRoy:2019hij,Sharma:2021kqb} as a two parameter ($n$ and $r_0$) family of simple Lorentzian wormholes, where $n$ is a free even-integer exponent and $r_0$ is the throat radius. This wormhole presents a shape function given by
\begin{equation}\label{berre}
b(r)=  r-r^{(3-2n)} (r^n-r_0^n)^{(2-\frac{2}{n})}.
\end{equation}
The exponent $n=2$ furnishes the original and simplest solution for that wormhole, in which $b(r)=r_0^2/r$.

The $f(R)$ model that we will henceforth employ is $f(R)=R+\alpha R^{m}$, which has a cosmological motivation, namely the pure Starobinsky one ($m=2$). It was shown that models with other values for $m$ (specifically, $1<m\leq 2$) can in addition cure the formation of curvature singularities in the gravitational collapse \cite{Bamba}.

   Considering that $R=2b'(r)/r^2$ and $b(r)$ is given by (\ref{berre}), we have

\begin{equation}\label{F(r)}
F(r)=1+2^{m-1}\alpha m\left[\frac{1}{r^{2}}+\frac{1}{r^{2n}}(2n-3)\left(r^{n}-r_{0}^{n}\right)^{2-\frac{2}{n}}+\frac{2}{r^{n}}(1-n)\left(r^{n}-r_{0}^{n}\right)^{1-\frac{2}{n}}\right]^{m-1}.
\end{equation}

Finally, let us define the dark matter content that we will use in our analysis. As in \cite{Jusufi:2019knb,Xu:2020wfm}, we will adopt simple but well-motivated dark matter density profiles with respect to the $\Lambda$CDM scenario. Compilations of observations from galaxies and their agglomerates, supported by numerical simulations \cite{Dubinski:1991bm,Navarro:1995iw,Navarro:1996gj}, suggest that the dark matter halo can be described by the Navarro-Frenk-White (NFW) density profile, whose analytical expression is defined as
\begin{equation}\label{NFW}
\rho=\rho_{NFW}=\frac{\rho_s}{\frac{r}{R_s}\left(1+\frac{r}{R_s}\right)^2},
\end{equation}
where $\rho_{s}$ is the density within the central region dominated by dark matter and $R_{s}$ is its characteristic radius. The NFW density profile accounts for a large family of models for dark matter from which the collision effects between particles are very weak.

The second case analyzed is based on the Bose-Einstein Condensation dark matter model and can be described by the Thomas-Fermi (TF) profile, whose application is more relevant in small distances of galaxies. In fact, the interactions of dark matter particles can no longer be neglected in the inner regions of galaxies, so it ceases to be cold. The TF profile can be represented in the form \cite{Boehmer:2007um}
\begin{equation}\label{TF}
\rho=\rho_{TF}=\rho_{s}\frac{\sin(k r)}{kr},
\end{equation}with $k=\pi/R_{s}$. The last case to be considered is the Pseudo isothermal (PI) profile, related to a class of dark matter models present in theories of modified gravity known as MOND (modified Newtonian dynamics) \cite{Begeman:1991iy}. The PI profile is given by
\begin{equation}\label{PI}
\rho=\rho_{PI}=\frac{\rho_s}{1+\left(\frac{r}{R_s}\right)^2}.
\end{equation}

\begin{figure}[h!]
    \centering
            \includegraphics[width=0.5\textwidth]{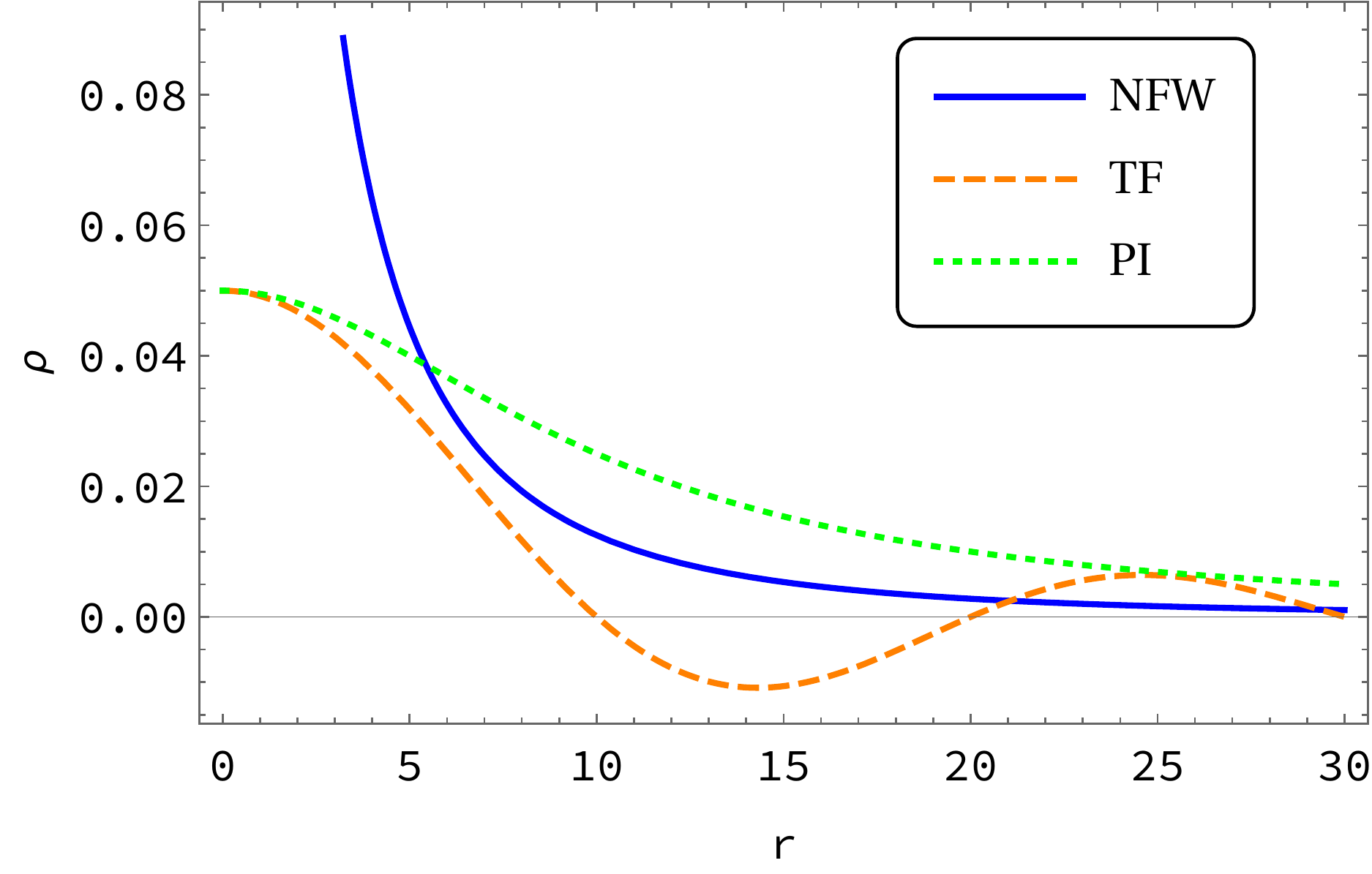}
        \caption{The behavior of dark matter density $\rho$ as a function of radial distance $r$ for the three analyzed profiles. The parameter settings are: $\rho_{s}=0.05$ and $R_{s}=10$ in Planck units.}
    \label{DarkDensities}
\end{figure}

In the context of wormhole studies, these profiles of dark matter density are examined in \cite{Jusufi:2019knb,Xu:2020wfm} and are depicted here in Fig. \ref{DarkDensities} as functions of the radial coordinate, $r$. The NFW model profile seems to grow indefinitely near the origin $r=0$. However, this feature proceeds from the relatively low resolution of the used numerical simulations (order of 1 kpc). Despite this, the radial coordinate is always greater than  (or equals to) the wormhole throat radius, $r_0$, and therefore the origin is out of our analysis. Also, Fig. \ref{DarkDensities} reveals that close to the wormhole throat, the dark matter density is positive for all cases and very similar in the TF and PI models, while the NFW profile is higher than the others.

\section{Energy conditions\label{EC}}

In order to investigate the energy conditions associated with the anisotropic dark matter as a source for the wormhole, we replace Eq. (\ref{F(r)}) and its derivatives, as well as the dark matter density profiles into Eqs. (\ref{Pe}) and (\ref{pezinho}). Thus, we find the expressions for the radial and lateral pressures, respectively. As these are quite involved, we will discuss the energy conditions graphically. Before this, we must notice that, by Eqs. (\ref{Pe})-(\ref{pezinho}), the Null Energy Conditions (NEC) and the Weak Energy Conditions (WEC), for which $\rho+p_i\geq 0$ (also $\rho\geq 0$ for the latter), are independent of the employed dark matter model. On the other hand, the more restrictive of the energy conditions (Strong Energy Conditions-SEC, $\rho+p_i\geq 0$, $\rho+\sum_i p_i\geq 0$) is model-dependent.
\begin{figure}[h!]
    \centering
    \begin{minipage}{0.5\textwidth}  \nonumber
        \centering
        \includegraphics[width=0.9\textwidth]{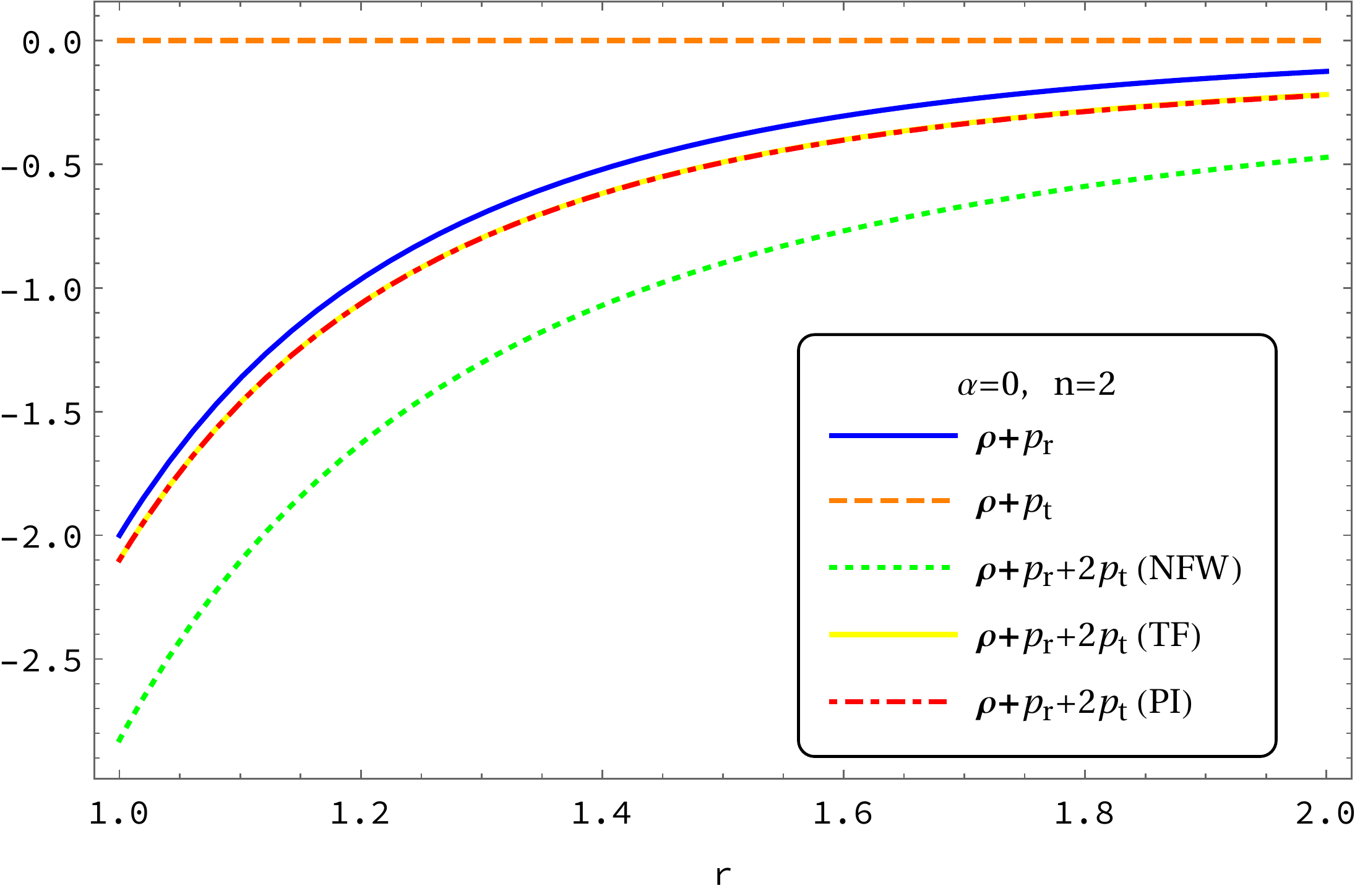}
    \end{minipage}\hfill
    \begin{minipage}{0.5\textwidth}  \nonumber
        \centering
        \includegraphics[width=0.9\textwidth]{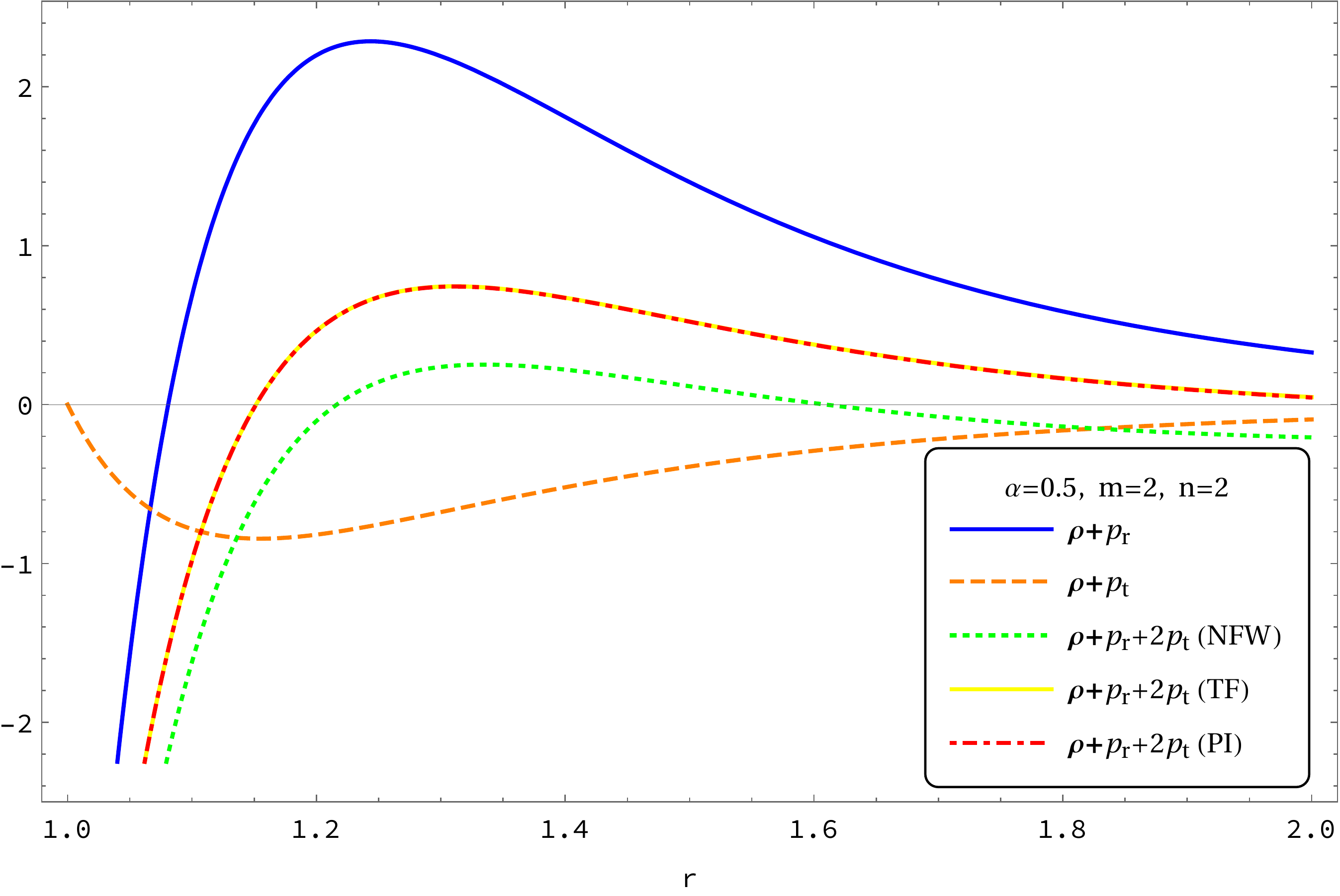}
    \end{minipage}\hfill
    \begin{minipage}{0.5\textwidth}  \nonumber
        \centering
        \includegraphics[width=0.9\textwidth]{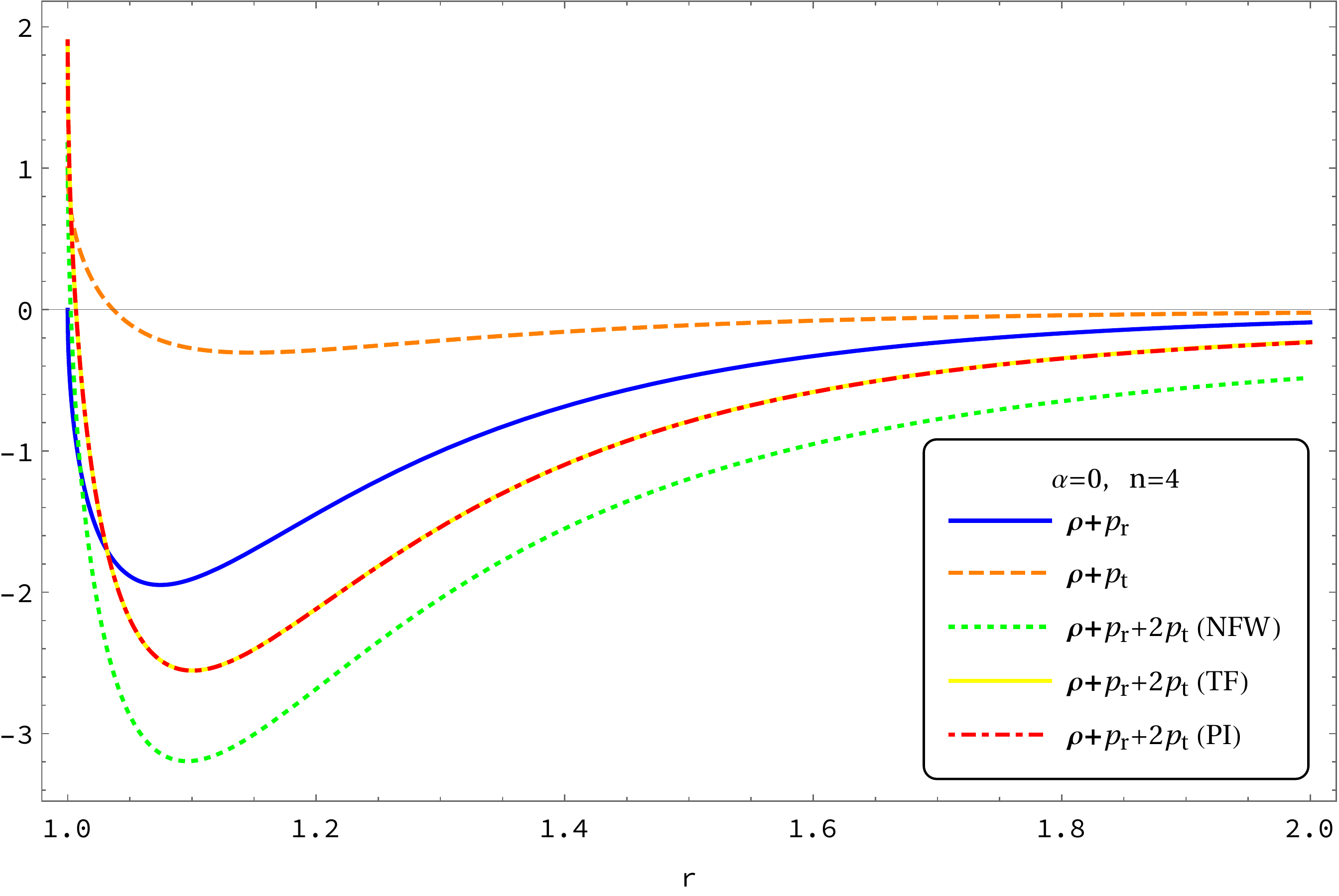}
    \end{minipage}\hfill
    \begin{minipage}{0.5\textwidth} \nonumber
        \centering
        \includegraphics[width=0.9\textwidth]{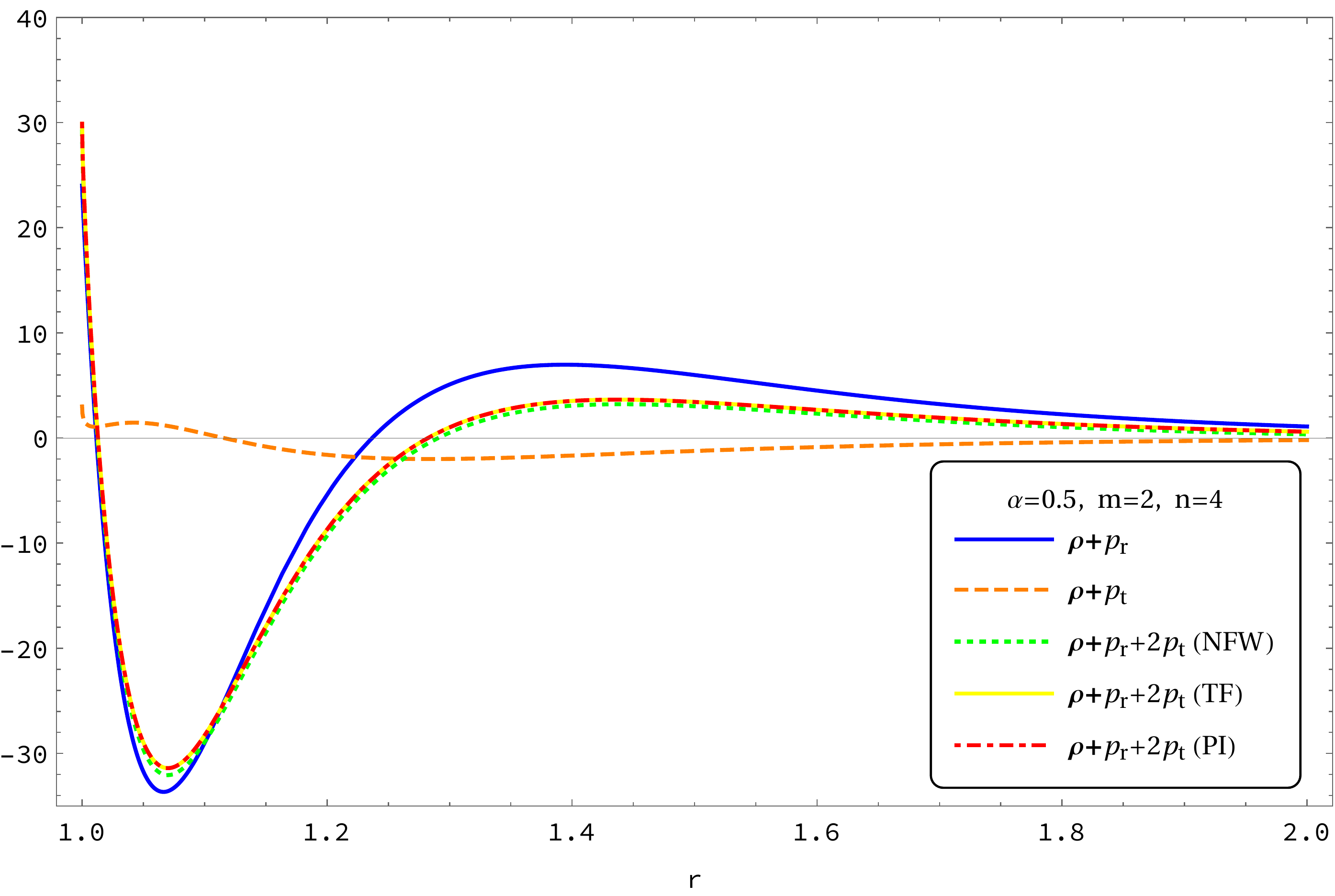}
    \end{minipage}
    \caption{Radial dependence of the energy density and pressure combinations for $n=2$ (top) and $n=4$ (bottom) Ellis-Bronikov wormhole solutions, in the pure Starobinsky $f(R)$ model ($m=2$) and for the three dark matter profiles, considering $\alpha=0$ (left panel) and $\alpha=0.5$ (right panel), $\rho_{s}=0.05$, $R_{s}=10$, and $r_0=1$, in Planck units.}
    \label{energy_conditions 1}
\end{figure}

In Fig. \ref{energy_conditions 1}, we depict the energy density and pressure combinations as functions of the radial coordinate in order to examine NEC, WEC, and SEC, considering that $\rho>0$. Notice that for $n=2$ (the simplest Ellis-Bronikov solution) and $\alpha=0$ ({\it i.e.}, in general relativity), none of these conditions are obeyed in all space ($r\geq r_0$), or they are partially satisfied if we observe that $\rho+p_t=0$. For $n\geq 4$, they are all met exactly at the wormhole throat. However, as  we can see through the general expressions valid at this point, namely
\begin{equation}\label{atthroat}
\left.(\rho+p_{r}+2p_{t})\right|_{r\rightarrow r_{0}}=\left\{ \begin{array}{cc}
2^{m}m(3-2m)\alpha\left(-\frac{1}{r_{0}^{2}}\right)^{m}-\frac{2}{r_{0}^{2}}\left(1+\frac{\rho_{s}r_{0}R_{s}^{3}}{\left(r_{0}+R_{s}\right){}^{2}}\right), & \ \ n=2;\\
2^{m}m(-5+6m)\alpha\left(\frac{1}{r_{0}^{2}}\right)^{m}+\frac{2}{r_{0}^{2}}\left(1-\frac{\rho_{s}r_{0}R_{s}^{3}}{\left(r_{0}+R_{s}\right){}^{2}}\right), & \ \ n=4;\\
2^{m}m\alpha\left(\frac{1}{r_{0}^{2}}\right)^{m}+\frac{2}{r_{0}^{2}}\left(1-\frac{\rho_{s}r_{0}R_{s}^{3}}{\left(r_{0}+R_{s}\right){}^{2}}\right), & \ \ n\geq6,
\end{array}\right.
\end{equation}
for $\alpha=0$, a greater concentration of dark matter may make it hard to fulfill SEC, since the total sum of the pressures with the density becomes negative when $\rho_s> (r_0+R_s)^2/(r_0R_s^3)$.
\begin{figure}[h!]
    \centering
    \begin{minipage}{0.5\textwidth}  \nonumber
        \centering
        \includegraphics[width=0.9\textwidth]{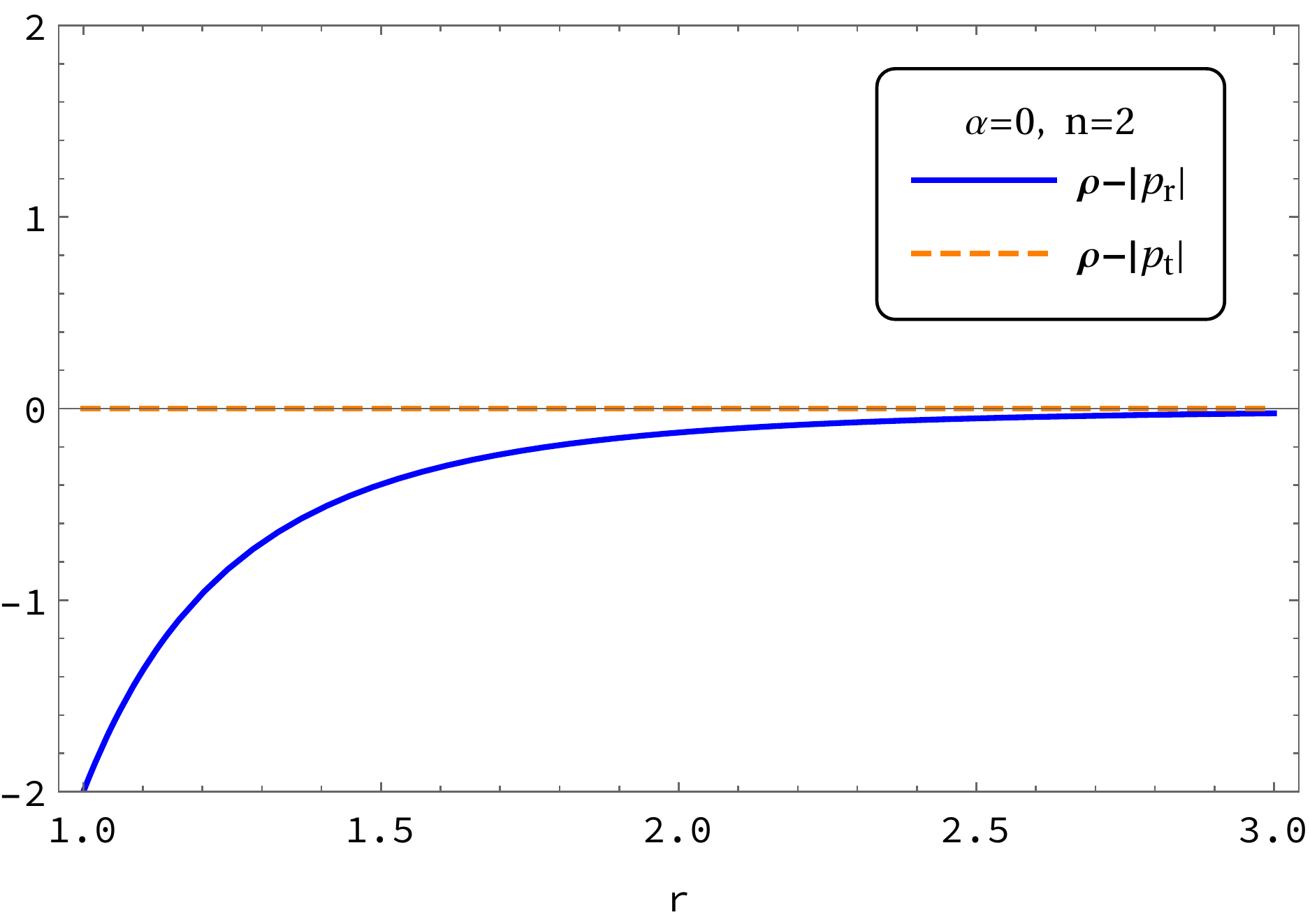}
    \end{minipage}\hfill
    \begin{minipage}{0.5\textwidth}  \nonumber
        \centering
        \includegraphics[width=0.9\textwidth]{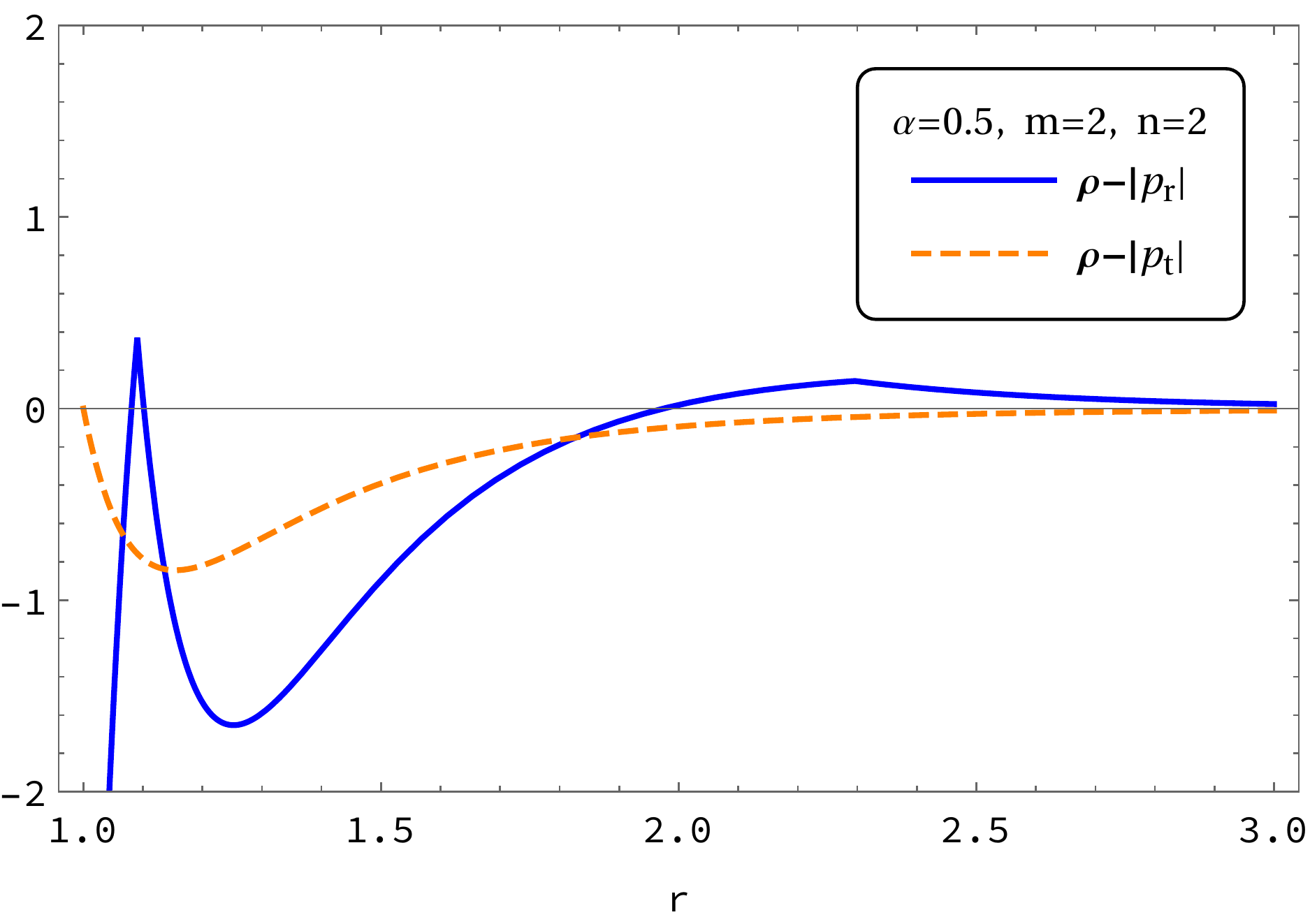}
    \end{minipage}\hfill
    \begin{minipage}{0.5\textwidth}  \nonumber
        \centering
        \includegraphics[width=0.9\textwidth]{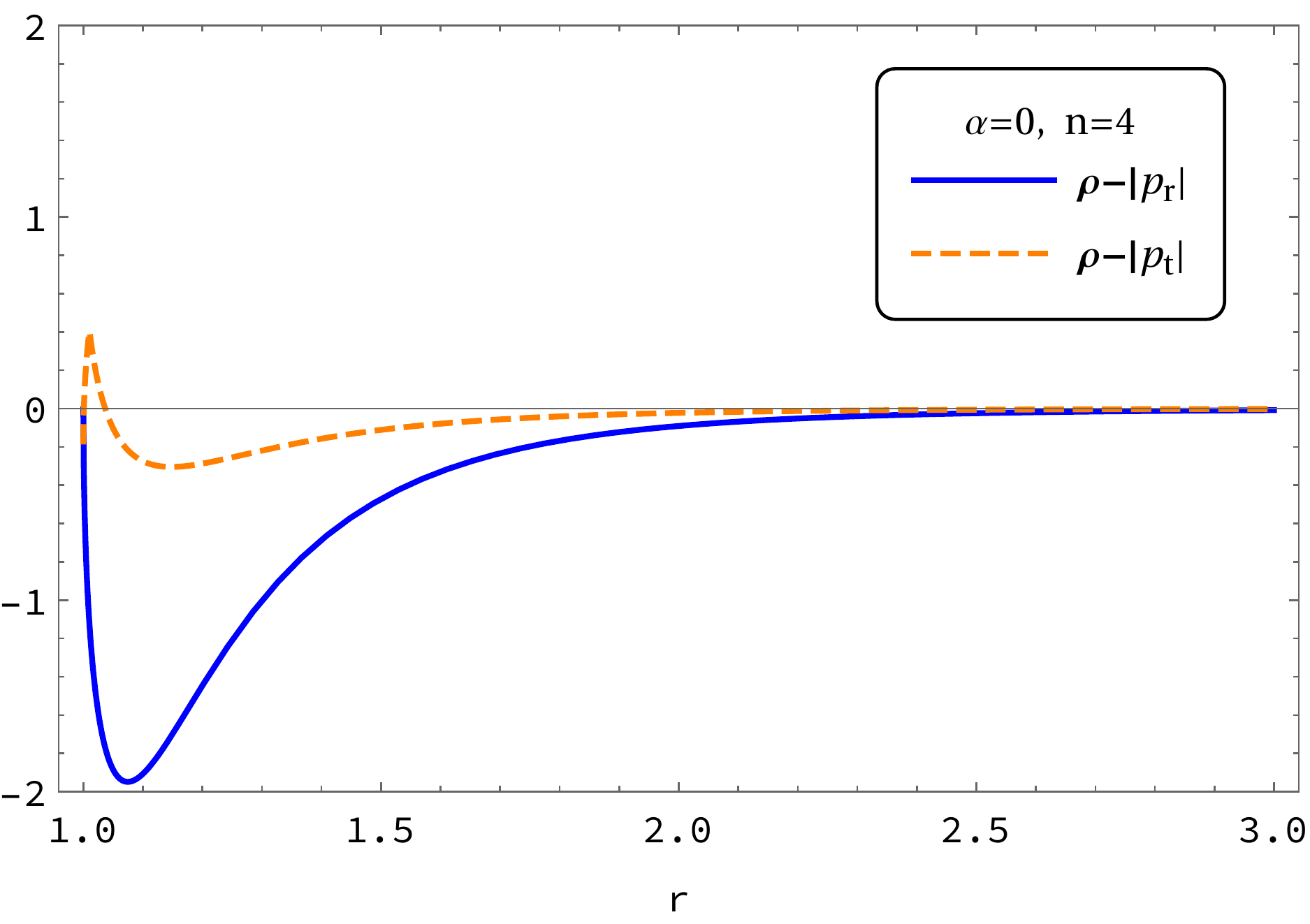}
    \end{minipage}\hfill
    \begin{minipage}{0.5\textwidth} \nonumber
        \centering
        \includegraphics[width=0.9\textwidth]{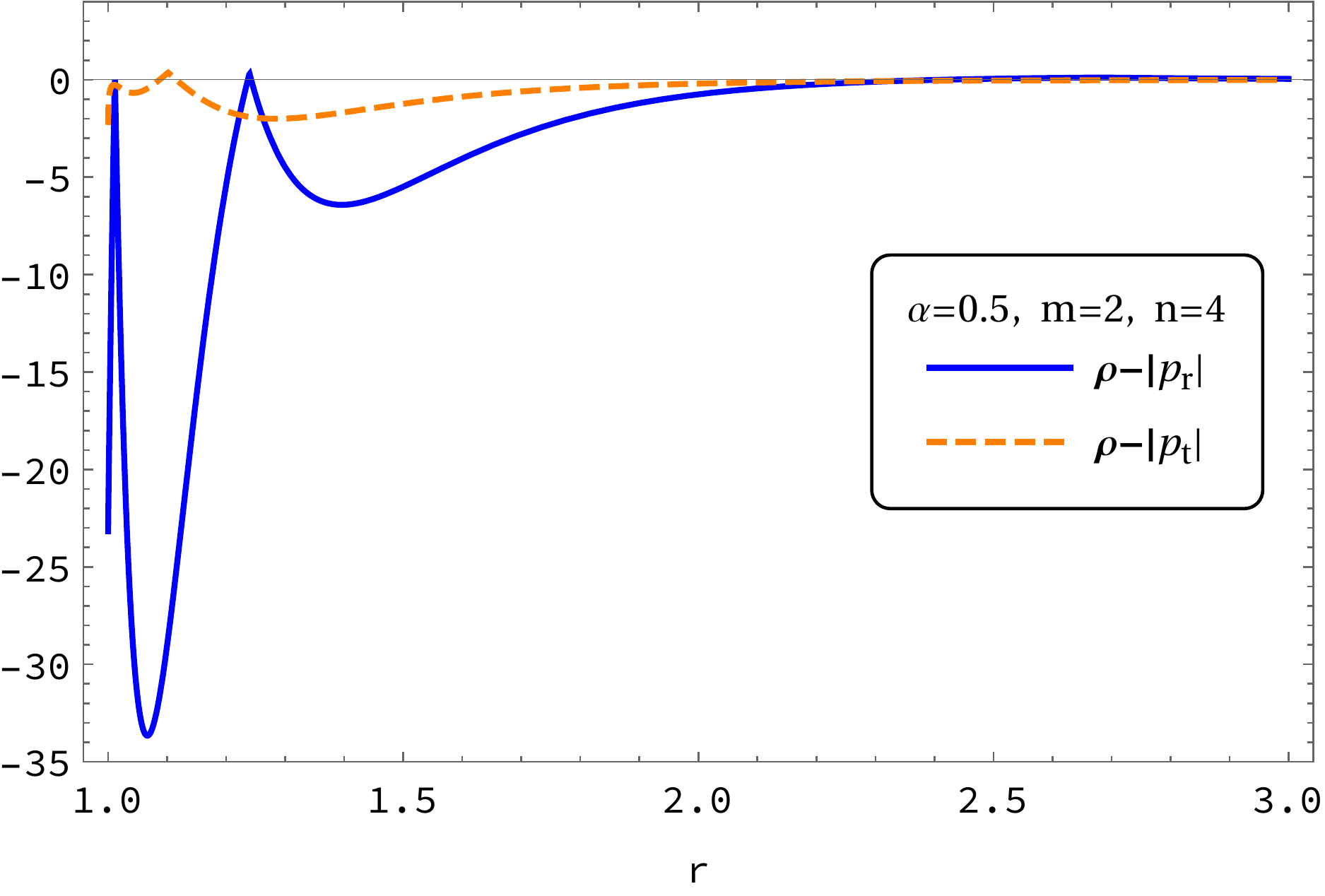}
    \end{minipage}
    \caption{The dominant energy condition (DEC): Radial dependence of the energy density and pressure $p_{r}$ and $p_{t}$ for $n=2$ (top) and $n=4$ (bottom) Ellis-Bronikov wormhole solutions, in the simplified Starobinsky $f(R)$ model ($m=2$) and NFW model for dark matter, considering $\alpha=0$ (left panel) and $\alpha=0.5$ (right panel), $\rho_{s}=0.05$, $R_{s}=10$, and $r_0=1$, in Planck units.}
    \label{DEC}
\end{figure}
\begin{figure}[h!]
    \centering
            \includegraphics[width=0.45\textwidth]{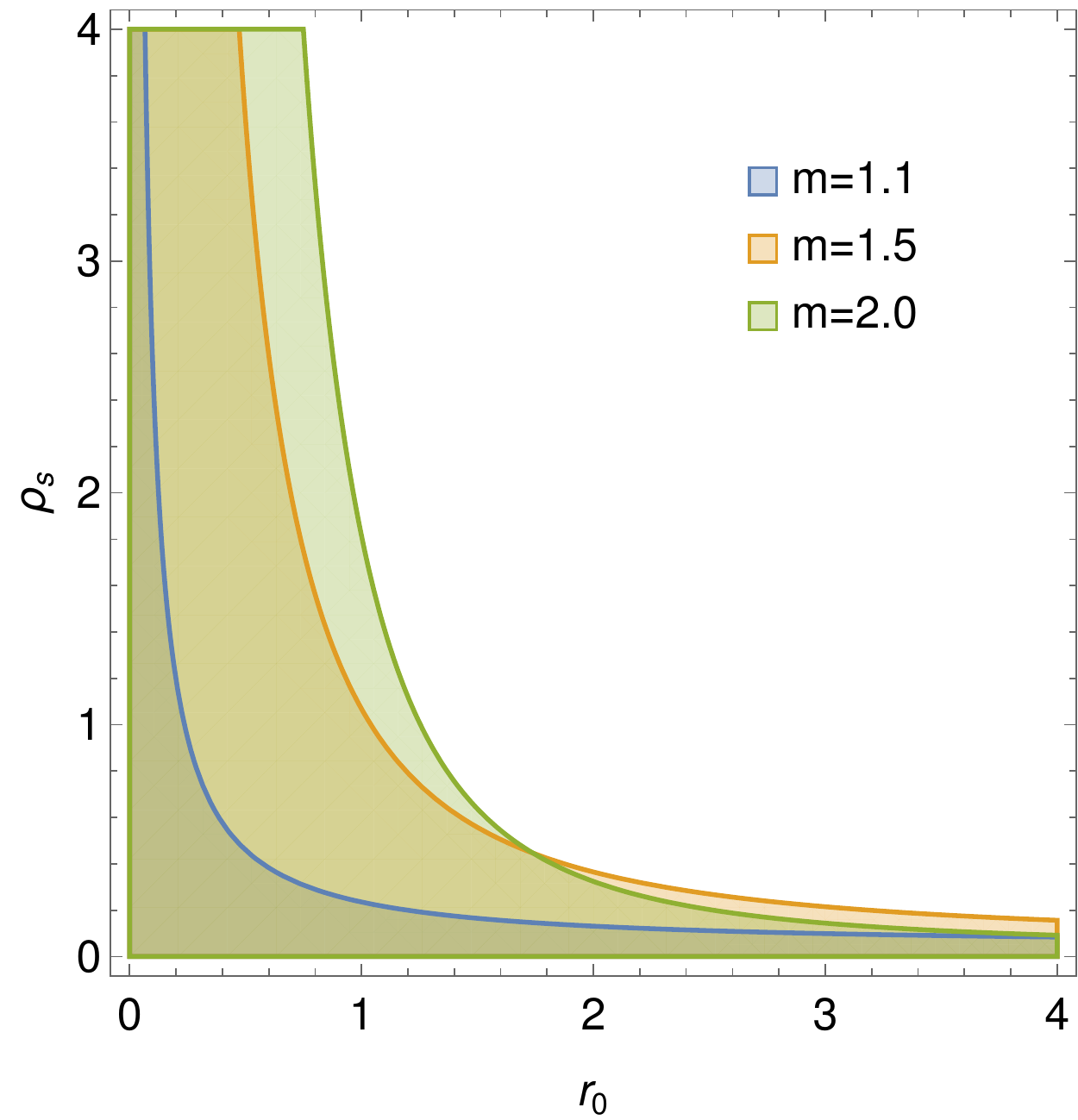}
        \caption{Parameter space where the regions with values of both the throat radius and dark matter characteristic density are highlighted, which allow SEC to be fulfilled at the wormhole throat, for some values of $m$, with $n=4$, $\alpha=0.8$, $R_s=9$, in NFW DM model (Planck units are used).}
    \label{ParameterSpace}
\end{figure}

Moving away from general relativity ({\it i.e.}, $\alpha\neq 0$; this constant must be positive in order to support cosmological observations \cite{Kehagias:2013mya}), in considering $n=2$, NEC, WEC and SEC are not met nearby and at the wormhole throat; however, SEC is partially obeyed in a limited region of the space sufficiently far from it. For $n=4$, those conditions are all satisfied nearby the throat, provided the content of dark matter is not much high. On the other hand, according to the second one of Eqs. (\ref{atthroat}), a greater coupling constant $\alpha$ can compensate for the growth of the dark matter and thus SEC to continue being satisfied at the throat. This conclusion is also valid for the solutions in which $n\geq 6$. Away from the throat, the present energy conditions are partially fulfilled for all the cases.

Although the three dark matter models under consideration are used in different contexts, they present the same properties in the wormhole formation, as evidenced in the previous analysis of the energy conditions, that is, the influence of different DM density profiles is very similar. As we can see, some associated behaviors practically coincide. Therefore, we will focus on the analysis of the NFW profile in what follows.

Regarding the Dominant Energy Conditions (DEC), in which $-\rho\leq p_i \leq \rho$, Fig. (\ref{DEC}) depicts $\rho-|p_i|$ as functions of the radial coordinate, $r$, for diverse situations. Considering $\alpha=0$ and $n=2$, those conditions are partially met with respect to the lateral pressure, in all space. For $n\geq 4$, DEC is not satisfied anywhere. When $\alpha\neq 0$, the energy conditions under consideration are partially satisfied in a few regions of space, for $n\geq 2$.

In Fig. \ref{ParameterSpace} we depict the parameter space where, in the colored region, $\rho+p_r+2p_t\geq 0$ is represented at the wormhole throat, for some values of $m$. Notice that the parameter space is the largest with respect to the Starobinsky model ($m=2$) for small throat radius and high DM concentrations. In other words, it is the most energetically favorable. However, the other power-law $f(R)$ models ($m<2$) are the most favored when one considers greater throat radius and low DM densities. Therefore, from the point of view of fulfillment of SEC, the model with the greater deviation from GR prevails over the other in scenarios where one has very high DM densities and microscopic wormholes, as expected.

\section{Conclusions\label{Con}}

We have studied the generalized Ellis-Bronikov (E-B) traversable wormhole solutions, which depend on a free parameter besides the throat radius, in $f(R)$ extended gravity. The shape function associated with the wormhole was therefore fixed. Because the object is zero-tidal, we can only consider an anisotropic dark matter as a source for it. We have analyzed three DM phenomenological models, namely, those of Navarro-Frenk-White (NFW), Thomas-Fermi (T-F), and Pseudo-isothermal (P-I). Hence, we found the correct field equations by particularizing the approach to the Starobinky-like $f(R)$ power-law model.

Following, we have analyzed the energy conditions, focusing on SEC and DEC, since they depend on the material source profile, differently from that occurs with NEC and WEC. Dark matter does not obey SEC anywhere in GR (where the coupling constant $\alpha$ vanishes) for the simplest E-B wormhole ($n=2$). These energy conditions are precisely met at the wormhole throat in the other E-B models ($n\geq 4$). Nevertheless, a higher amount of dark matter can undermine this fulfillment. Concerning DEC, it is not obeyed anywhere in all situations.

In taking into account the quadratic gravity, for $n=2$ E-B wormhole, SEC is partially fulfilled in a limited region of space. On the other hand, these energy conditions are entirely fulfilled nearby and at the wormhole throat for $n\geq4$. On the other hand, a higher concentration of dark matter can spoil such fulfillment, as in RG case. But a greater coupling constant can restore it at the wormhole throat, as well as a smaller size of this latter. Fig. \ref{ParameterSpace} illustrates such competition between the involved parameters and some power-law $f(R)$ models, being the pure Starobinsky model ($m=2$) the most energetically favorable for small throat radii and high concentration of DM since the associated parameter space is the largest. Otherwise, the power-law $f(R)$ models with smaller exponents are the most favored. Still, in this context, we have shown that, unlike GR, DEC is partially satisfied in limited regions of space for all $n$.

We have thus shown that different models of DM  present the same properties in the wormhole formation. Finally, we can conclude that anisotropic dark matter can support a class of traversable wormholes as non-exotic matter, at least in regions nearby and at the throat, in the simplest phenomenological $f(R)$ extended theory of gravity.

\acknowledgments{The authors thank the Conselho Nacional de Desenvolvimento Cient\'{i}fico e Tecnol\'{o}ogico (CNPq), grants no 308268/2021-6 (CRM) and no 311732/2021-6 (RVM) for financial support.}

 \end{document}